\documentclass[conference,a4paper]{IEEEtran}
\usepackage{amsmath, amsthm, amssymb}
\usepackage{graphicx} 
 \usepackage{amsthm}
\newtheorem{lemma}{Lemma}

\newtheorem{theorem}{Theorem}

\usepackage{color}

\usepackage{rotating,cite}

\usepackage{epstopdf}

\newcommand{\bfm}[1]{\mbox{\boldmath $#1$}}

\begin{document}
\sloppy
\title{An Upper Bound on the Capacity of non-Binary Deletion Channels\vspace*{-.15in}}

\author{\IEEEauthorblockN{Mojtaba Rahmati}\\
\vspace*{-.1in}
\IEEEauthorblockA{School of Electrical, Computer and Energy Engineering\\
Arizona State University, Tempe, AZ 85287--5706, USA\\
Email: mojtaba@asu.edu\vspace*{-.38in}}   \and \IEEEauthorblockN{Tolga M. Duman}\\
\vspace*{-.1in}
\IEEEauthorblockA{Department of Electrical and Electronics Engineering\\ Bilkent University, Bilkent, Ankara, 06800, Turkey\\
Email: duman@ee.bilkent.edu.tr}
\vspace*{-.2in}}

\maketitle

\begin{abstract}
\let\thefootnote\relax\footnotetext{T. M. Duman is currently with Bilkent University in Turkey, and on leave from Arizona State University, Tempe, AZ.} 
We derive an upper bound on the capacity of non-binary deletion channels. Although binary deletion channels have received significant attention over the years, and many upper and lower bounds on their capacity have been derived, such studies for the non-binary case are largely missing. The state of the art is the following: as a trivial upper bound, capacity of an erasure channel with the same input alphabet as the deletion channel can be used, and as a lower bound the results by Diggavi and Grossglauser in~\cite{diggavi2006information} are available. In this paper, we derive the first non-trivial non-binary deletion channel capacity upper bound and reduce the gap with the existing achievable rates. To derive the results we first prove an inequality between the capacity of a $2K$-ary deletion channel with deletion probability $d$, denoted by $C_{2K}(d)$, and the capacity of the binary deletion channel with the same deletion probability, $C_2(d)$, that is, $C_{2K}(d)\leq C_2(d)+(1-d)\log(K)$. Then by employing some existing upper bounds on the capacity of the binary deletion channel, we obtain upper bounds on the capacity of the $2K$-ary deletion channel. We illustrate via examples the use of the new bounds and discuss their asymptotic behavior as $d \rightarrow 0$. 
\end{abstract}

\section{Introduction}
Non-binary deletion channels can be used to model information transmission over a finite buffer channel~\cite{diggavi2006information}, where a packet (non-binary symbol) loss occurs whenever a packet arrives at a full buffer. When the channel drop-outs are independent and identically distributed (i.i.d.), the channel is referred as a non-binary i.i.d. deletion channel. Dobrushin~\cite{dobrushin} proved the existence of Shannon's theorem for discrete memoryless channels with synchronization errors. As a result, Shannon's theorem holds in non-binary deletion channels and information and transmission capacities are equal. 

In this paper, we focus on a $2K$-ary deletion channel in which every transmitted symbol is either lost through the transmission with probability of $d$ or received correctly with probability of $1-d$. There is no information about the position of the lost symbols at either the transmitter or the receiver. Clearly the capacity of a $2K$-ary erasure channel with erasure probability $d$ is an upper bound on the capacity of the $2K$-ary deletion channel since by revealing information about the position of the lost symbols to the receiver, the corresponding genie-aided deletion channel is nothing but an erasure channel. Therefore, for the capacity of the $2K$-ary input deletion channel $C_{2K}(d)$, the relation $C_{2K}(d)\leq (1-d)\log(2K)$ holds. Besides this trivial upper bound, to the best of our knowledge, there are no other (tighter) upper bounds on the capacity of non-binary deletion channels. 

Our main result is to relate the capacity of a $2K$-ary deletion channel with deletion probability $d$ to the capacity of the binary deletion channel with deletion probability $d$ by the inequality $C_{2K}\leq C_2(d)+(1-d)\log(K)$. As a result, any upper bound on the binary deletion channel capacity can be used to derive an upper bound on the $2K$-ary deletion channel capacity. For example, by using the result from~\cite{IT-quasi}, we obtain $C_{2K}(d)\leq (\log(K)+0.4143)(1-d)$ for $d\geq 0.65$.

The paper is organized as follows. In Section~\ref{sec-previous-work}, we briefly review the existing work on the capacity of binary and non-binary deletion channels. In Section~\ref{sec-Preliminaries}, we first give the general $2K$-ary deletion channel model and then we observe that it can be considered as a parallel concatenation of $K$ independent deletion channels (where each input is binary). Also in the same section, we discuss the possible generalization of 
the existing Blahut-Arimoto algorithm (BAA) based upper bounding approaches (useful for the binary deletion channels) to the case of $2K$-ary deletion channels. In Section~\ref{sec-Proof-Thm}, we prove the main result of the paper providing an upper bound on $C_{2K}(d)$ in terms of $C_2(d)$. In Section~\ref{sec-Numerical-Examples}, several implications of the result are given where we compare the resulting capacity upper bounds with the existing capacity upper and lower bounds, and we provide a discussion of the channel capacity behavior as the deletion probability approaches zero. Finally, we conclude the paper in Section~\ref{sec-Conclusions}.

\section{Previous Works}\label{sec-previous-work}

Capacity of binary deletion channels has received significant attention in the existing literature, e.g., see~\cite{mitzenmacher-survey} and references therein. There are several results on capacity lower bounds~\cite{gallager,drinea2006lower, drinea2007}. Gallager~\cite{gallager} provided the first lower bound on the transmission capacity of the channels with random insertion, deletion and substitution errors which provides a lower bound on the binary deletion channel capacity as well. The tightest lower bound on the binary deletion channel capacity is provided in~\cite{drinea2007} where the information capacity of the binary deletion channel is directly lower bounded by considering input sequences as alternating blocks of zeros and ones (runs) and the length of the runs $L$ as i.i.d. random variables following a particular distribution over positive integers with a finite expectation and finite entropy.

There are also several upper bounds on the binary deletion channel capacity, e.g.,~\cite{diggavi-capacity,dario,IT-quasi}. In~\cite{diggavi-capacity} a genie-aided channel is considered in which the receiver is provided by side information about the completely deleted runs, e.g., in transmitting $``110001"$ over the original channel by deleting the entire run of zeros, the sequence $``111"$ is received while in the considered genie-aided channel $``11-1"$ represents the received sequence. Then an upper bound on the capacity per unit cost of the genie-aided channel is computed by running the BAA algorithm. Fertonani and Duman~\cite{dario}, by considering several different genie-aided channels, are able to derive tighter upper bounds on the binary deletion channel capacity compared to the results in~\cite{diggavi-capacity} for $d>0.05$. In~\cite{IT-quasi}, authors improve upon the upper bounds provided in~\cite{dario} for $d>0.65$ where they first derive an inequality relation among the capacity of three different binary deletion channels and as a special case they obtain $C_2(\lambda d+1-\lambda)\leq \lambda C_2(d)$ which shows that $C_2(d)\leq 0.4143(1-d)$ for $d\geq 0.65$.

To the best of our knowledge, the only non-trivial lower bounds on the capacity of the non-binary deletion channels are provided in~\cite{diggavi2006information} where two different bounds are derived. More precisely, the achievable rates of the $2K$-ary input deletion channel are computed for i.i.d. and Markovian codebooks by considering a simple decoder which decides in favor of a sequence if the received sequence is a subsequence of only one transmitted sequence. The derived achievable rates are given by
\begin{equation}\label{eq-LB-diggavi-iid}
C_{2K}\geq \log\left(\frac{2K}{2K-1}\right)+(1-d)\log(2K-1)-H_b(d),
\end{equation}
by considering i.i.d. codebooks, where $H_b(d)=-d\log(d)-(1-d)\log(1-d)$, and
\begin{equation}\label{eq-LB-diggavi-Markov}
C_{2K}\hspace*{-.03in}\geq\hspace*{-.05in}  \sup_{\gamma>0,\ 0<p<1}\hspace*{-.03in}[-(1\hspace*{-.02in}-\hspace*{-.02in}d)\log\left((1\hspace*{-.02in}-\hspace*{-.02in}q)A\hspace*{-.02in}+\hspace*{-.02in}qB\right)\hspace*{-.02in}-\hspace*{-.02in}\gamma\log(e)]
\end{equation}
by considering Markovian codebooks, with $q=\frac{1}{2K}\left(\hspace*{-.02in}1\hspace*{-.02in}+\hspace*{-.02in}\frac{(1-d)(2K-1)(2Kp-1)}{2K-1-d(2Kp-1)}\hspace*{-.02in}\right)$, $A\hspace{-.03in}=\hspace{-.03in}\frac{e^{-\gamma}(1-p)}{(2K\hspace*{-.02in}-\hspace*{-.02in}1)(1-e^{-\gamma}(1-\frac{1-p}{2K-1}))}$ and $B\hspace*{-.03in} =\hspace*{-.03in}e^{-\gamma}\left((1-p)A\hspace*{-.02in}+\hspace*{-.02in}p\right)$. Non-binary input alphabet channels with synchronization errors are also considered in~\cite{mercier2012} where the capacity of memoryless synchronization error channels in the presence of noise and the capacity of channels with weak synchronization errors (i.e., the transmitter and receiver are partly synchronized) have been studied. The main focus of the work in~\cite{mercier2012} is on the asymptotic behavior of the channel capacity for large values of $K$.

\section{Preliminaries}\label{sec-Preliminaries}

\subsection{Channel Model}\label{sec-2K-channel-model}
An i.i.d. $2K$-ary input deletion channel with input alphabet ${\cal{X}}=\{1,\dotsc,2K\}$ is considered in which every transmitted symbol is either randomly deleted with probability $d$ or received correctly with probability $1-d$ while there is no information about the values or the position of the lost symbols at the transmitter and the receiver. In transmission of $N$ symbols through the channel, the input sequence is denoted by $\bfm X=(x_1,\dotsc,x_N)$ in which $x_n\in {\cal{X}}$ and $\bfm X\in {\cal{X}}^N$, and the output sequence is denoted by $\bfm Y=(y_1,\dotsc,y_M)$ in which $M$ is a binomial random variable with parameters $N$ and $d$ (due to the characteristics of the i.i.d. deletion channel).

\subsubsection*{A Different Look at the $2K$-ary Deletion Channel}\label{sec-new-defined-channel}
\begin{figure}[t]
\centerline{
  \hbox{
    {\includegraphics[trim=28mm 70mm 15mm 46mm,clip,  width=.5\textwidth]{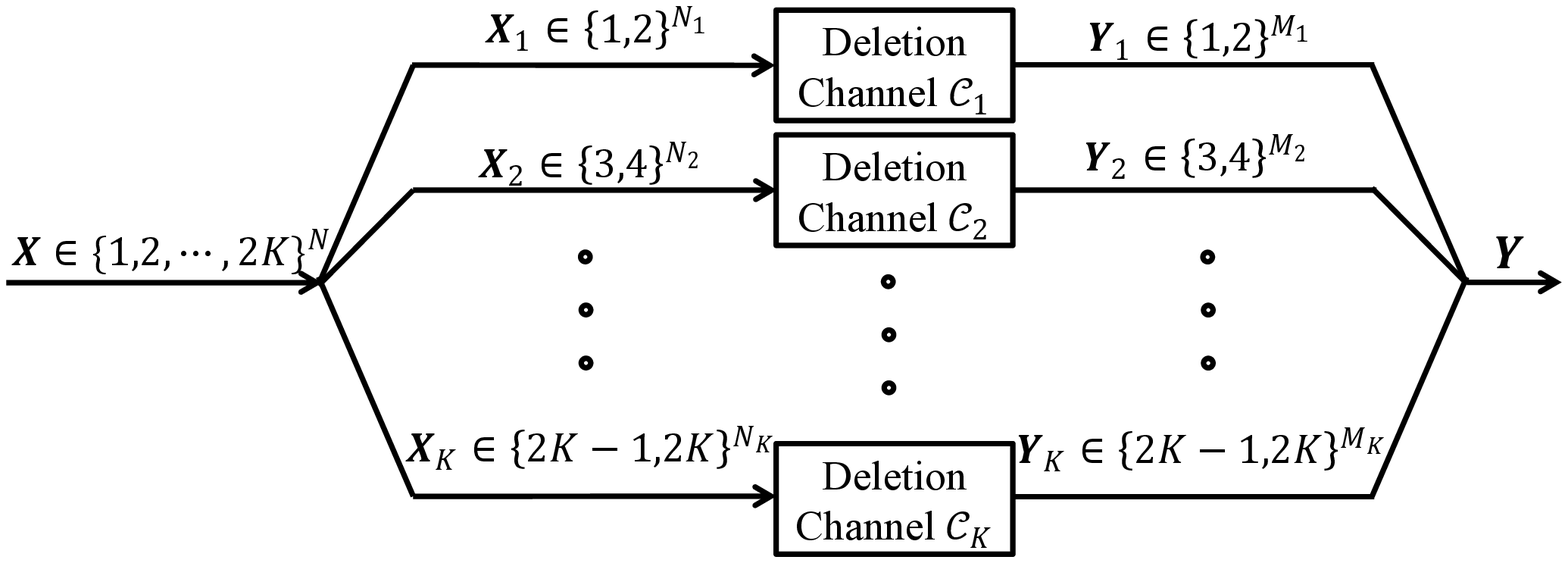}}
  }
}
\vspace*{-.1in}\caption{$2K$-ary deletion channel as a parallel concatenation of $K$ independent binary input deletion channels.\vspace*{-.15in}}
\label{fig-ch_model_2Kary}
\end{figure}

Any $2K$-ary input deletion channel with deletion probability $d$ can be considered as a parallel concatenation of $K$ independent binary deletion channels ${\cal{C}}_k$ ($k\in\{1,\dotsc,K\}$) all with the same deletion probability $d$, as shown in Fig.~\ref{fig-ch_model_2Kary}, in which the input symbols $2k-1$ and $2k$ travel through ${\cal{C}}_k$ and the surviving output symbols of the subchannels are combined based on the order in which they go through the subchannels. $\bfm X_k$ and $\bfm Y_k$ denote the input and output sequences of the $k$-th channel, respectively, and $N_k$ and $M_k$ denote the length of $\bfm X_k$ and $\bfm Y_k$, respectively.

To be able to relate the mutual information between the input and output sequences of the $2K$-ary deletion channel, $I(\bfm X;\bfm Y)$, with the mutual information between the input and output sequences of the considered binary deletion channels, $I(\bfm X_k;\bfm Y_k)$, we define two new random vectors $\bfm F_x=(f_x[1],\dotsc,f_x[N])$ and $\bfm F_y=(f_y[1],\dotsc,f_y[M])$ where $f_x[n]\in\{1,\dotsc,K\}$ and $f_y[m]\in\{1,\dotsc,K\}$ denote the label of the subchannel the $n$-th input symbol and $m$-th output symbol belong to, respectively. Clearly, by knowing $\bfm X$, one can determine ($\bfm X_1,\dotsc,\bfm X_K,\bfm F_x$) and by knowing ($\bfm X_1,\dotsc,\bfm X_K,\bfm F_x$) can determine $\bfm X$. The same situation holds for $\bfm Y$ and ($\bfm Y_1,\dotsc,\bfm Y_K,\bfm F_y$). Therefore, we have 
\vspace*{-.03in}\begin{align}\label{eq-I_2K}
I(\bfm X;\bfm Y)&=I(\bfm X_1,\dotsc, \bfm X_K,\bfm F_x;\bfm Y_1,\dotsc, \bfm Y_K,\bfm F_y)\nonumber\\
&=\sum_{k=1}^K I_k+I_F,
\end{align}
where $I_k=I(\bfm X_1,\dotsc, \bfm X_K,\bfm F_x;\bfm Y_k|\bfm Y_1,\dotsc,\bfm Y_{k-1})$ and 
\begin{equation}\label{eq-I_F}
I_F=I(\bfm X_1,\dotsc,\bfm X_K,\bfm F_x;\bfm F_y|\bfm Y_1,\dotsc,\bfm Y_K).
\end{equation}
In Section~\ref{sec-Proof-Thm}, we will derive upper bounds on $I_k$ and $I_F$ which will enable us to 
relate the non-binary and binary deletion channels capacities, and will lead to the main result of the paper.

\subsection{Discussion on BAA Based Upper Bounds}

One approach to derive upper bounds on the $2K$-ary deletion channel capacity is to modify the numerical approaches in~\cite{diggavi-capacity, dario} in which the decoder (and possibly the encoder) of the deletion channel is provided with some side information about the deletion process and the capacity (or an upper bound on the capacity) of the resulting genie-aided channel is computed by the Blahut-Arimoto algorithm. Although this approach is useful for binary input channels (even when other impairments such as insertions and substitutions are considered~\cite{dario2}), for the non-binary case, running the BAA for large values of $K$ is not computationally feasible. E.g., one of the upper bounds in~\cite{dario} is obtained by computing the capacity of the binary deletion channel with finite length of transmission $L=17$. Obviously, by increasing the alphabet size, $2K$, the maximum possible value of $L$ in running the BAA algorithm decreases. Therefore, to achieve meaningful upper bounds, $L$ needs to be increased which makes the numerical computations infeasible.

The main contribution of the present paper is that we are able to relate the capacity of the $2K$-ary deletion channel to the binary deletion channel capacity through an inequality which enables us to upper bound the $2K$-ary deletion channel capacity avoiding computationally formidable BAA directly for the $2K$-ary deletion channel.

\section{A Novel Upper Bound on $C_{2K}(d)$}\label{sec-Proof-Thm}
As introduced in Section~\ref{sec-2K-channel-model}, a $2K$-ary deletion channel can be considered as a parallel concatenation of $K$ independent binary deletion channels. This new look at a $2K$-ary deletion channel enables us to relate the $2K$-ary deletion channel capacity to the binary deletion channel capacity with the same deletion error probability as given in the following theorem.
\begin{theorem}\label{thm-2K-2}
Let $C_{2K}(d)$ denote the capacity of a $2K$-ary i.i.d. deletion channel with deletion probability $d$, then
\begin{equation}\label{eq-C2K-C2}
C_{2K}(d)\leq C_{2}(d)+(1-d)\log(K).
\end{equation}
\end{theorem}

As given in~\eqref{eq-I_2K}, the mutual information $I(\bfm X;\bfm Y)$ can be expanded in terms of several other mutual information terms, $I_k$ for $k\hspace*{-.02in}\in\hspace*{-.02in}\{1,\hspace*{-.01in}\dotsc\hspace*{-.01in},K\hspace*{-.02in}\}$ and $I_{\hspace*{-.01in}F}$. To prove the theorem, we first derive upper bounds on $I_k$ and $I_F$ in the following two lemmas.

\begin{lemma}\label{lemma-Ik}
For any input distribution $P(\bfm X_1,\dotsc,\bfm X_K,\bfm F_x)$, the mutual information $I_k$ given in~\eqref{eq-I_2K} can be upper bounded by
\begin{equation}
I_k\leq E\{N_k\} C_2(d)+2\log(N+1),\nonumber
\end{equation}
where $E\{.\}$ denotes the expected value.
\end{lemma}
\begin{IEEEproof}
For $I_k$, since $P(\bfm Y_k|\bfm Y_1,\dotsc,\bfm Y_{k-1},\bfm X_k)=P(\bfm Y_k|\bfm X_k)$ and $P(\bfm Y_k|\bfm X_1,\dotsc,\bfm X_K,\bfm F_x,\bfm Y_1,\dotsc,\bfm Y_{k-1})=P(\bfm Y_k|\bfm X_k)$, we can write
\begin{align}\label{eq-I_k}
I_k=&\ I(\bfm X_k;\bfm Y_k|\bfm Y_1,\dotsc, \bfm Y_{k-1})\nonumber\\
+I(&\bfm X_{\hspace*{-.01in}1} , ... , \bfm X_{\hspace*{-.01in}k-1},\bfm X_{\hspace*{-.01in}k+1} , ... ,\bfm X_{\hspace*{-.01in}K} ,\bfm F_{\hspace*{-.01in}x};\bfm Y_{\hspace*{-.01in}k}|\bfm Y_{\hspace*{-.01in}1},... ,\bfm Y_{\hspace*{-.01in}k-1},\bfm X_{\hspace*{-.01in}k})\nonumber\\
=&\ I(\bfm X_k;\bfm Y_k|\bfm Y_1,\dotsc, \bfm Y_{k-1})\nonumber\\
=&\ H(\bfm Y_k|\bfm Y_1,\dotsc, \bfm Y_{k-1})-H(\bfm Y_k|\bfm Y_1,\dotsc, \bfm Y_{k-1},\bfm X_k)\nonumber\\
=&\ H(\bfm Y_k)-I(\bfm Y_1,\dotsc, \bfm Y_{k-1};\bfm Y_k)-H(\bfm Y_k|\bfm X_k)\nonumber\\
\leq& \ I(\bfm X_k;\bfm Y_k).
\end{align}
Furthermore, $I(\bfm X_k;\bfm Y_k)$ can be written as
\begin{align}
I(\bfm X_k;\bfm Y_k)=&I(\bfm X_k;\bfm Y_k,N_k)-I(\bfm X_k;N_k|\bfm Y_k)\nonumber\\
=&I(\bfm X_k;\bfm Y_k|N_k)\hspace*{-.008in}+\hspace*{-.008in}I(\bfm X_k;N_k)\hspace*{-.008in}-\hspace*{-.008in}I(\bfm X_k;N_k|\bfm Y_k).\nonumber
\end{align}
Since $H(N_k|\bfm X_k)=0$ and \mbox{$I(\bfm X_k;N_k|\bfm Y_k)\geq 0$}, we arrive at
\begin{align}\label{eq-IXkYK}
\hspace*{-.095in} I(\bfm X_k;\bfm Y_k)\leq & I(\bfm X_k;\bfm Y_k|N_k)+H(N_k)\nonumber\\
\leq & I(\bfm X_k;\bfm Y_k|N_k)+\log(N+1)\nonumber\\
=&\hspace*{-.05in}\sum_{n_k=0}^{N}\hspace*{-.07in}P(N_k\hspace*{-.03in}=\hspace*{-.03in}n_k)I(\bfm X_k\hspace*{-.015in};\hspace*{-.02in}\bfm Y_k|n_k)\hspace*{-.03in}+\hspace*{-.03in}\log(N+1),
\end{align}
where the second inequality results since there are $N+1$ possibilities for $N_k$ and as a result $H(N_k)\leq \log(N+1)$. Furthermore, as shown in~\cite{dario}, for a finite length transmission over the deletion channel, the mutual information rate between the transmitted and received sequences can be upper bounded in terms of the capacity of the channel after adding some appropriate term, which can be spelled out as \cite[Eqn.~(39)]{dario}
\begin{equation}\label{eq-dario}
I(\bfm X_k;\bfm Y_k|N_k=n_k)\leq n_kC_2(d)+H(\bfm D_k|N_k=n_k),
\end{equation} 
where $\bfm D_k$ denotes the number of deletions through the transmission of $N_k$ bits over the $k$-th channel. We have
\begin{eqnarray}\label{eq-HDkNk}
H(\bfm D_k|N_k=n_k)&=&-\sum_{n=0}^{n_k}P(n_k,n,d)\log\left(P(n_k,n,d)\right)\nonumber\\
&\leq& \log{(n_k+1)}\leq  \log{(N+1)},
\end{eqnarray}
with $\displaystyle P(n_k,n,d)={n_k\choose n}d^n(1-d)^{n_k-n}$. Substituting~\eqref{eq-HDkNk} and \eqref{eq-dario} into \eqref{eq-IXkYK}, we obtain
\begin{align}
I(\bfm X_k;\bfm Y_k)\leq&  \sum_{n_k=0}^{N}P(N_k=n_k)\left(n_kC_2(d)\right)+2\log(N+1)\nonumber\\
=&E\{N_k\}C_2(d)+2\log{(N+1)}.\nonumber
\end{align}
Finally, by substituting the above inequality into~\eqref{eq-I_k}, the proof follows.
\end{IEEEproof}

\begin{lemma}\label{lemma-IF}
For any input distribution, the mutual information $I_F$ given in~\eqref{eq-I_F} can be upper bounded by
$$I_F\leq N(1-d)\log(K).$$ 
\end{lemma}
\begin{IEEEproof}
Using the definition of the mutual information, we can write
\begin{align}
I_F\hspace*{-.03in}=&H(\bfm F_{\hspace*{-.02in}y}|\bfm Y_{\hspace*{-.03in}1},\dotsc,\hspace*{-.01in}\bfm Y_{\hspace*{-.03in}K})\hspace*{-.03in}-\hspace*{-.03in}H\hspace*{-.02in}(\bfm F_{\hspace*{-.02in}y}|\bfm Y_{\hspace*{-.03in}1},\dotsc,\hspace*{-.01in}\bfm Y_{\hspace*{-.03in}K},\bfm X_{\hspace*{-.03in}1},\dotsc,\bfm X_{\hspace*{-.03in}K},\hspace*{-.01in}\bfm F_{\hspace*{-.02in}x}\hspace*{-.02in})\nonumber\\
\leq &H(\bfm F_{\hspace*{-.02in}y}|\bfm Y_{\hspace*{-.03in}1},\dotsc,\bfm Y_{\hspace*{-.03in}K})\nonumber\\
\leq &H(\bfm F_{\hspace*{-.02in}y}|M_1,\dotsc,M_K),
\end{align}
where the last inequality follows since $(\hspace*{-.01in} M_{\hspace*{-.01in}1},\hspace*{-.01in}\dotsc\hspace*{-.01in},M_{\hspace*{-.01in}K}\hspace*{-.0in})$ is a function of $(\bfm Y_{\hspace*{-.01in}1},\dotsc,\bfm Y_{\hspace*{-.01in}K})$, i.e., $H(\hspace*{-.01in}M_{\hspace*{-.01in}1},\dotsc,M_{\hspace*{-.01in}K}|\bfm Y_{\hspace*{-.01in}1},\dotsc,\bfm Y_{\hspace*{-.01in}K}\hspace*{-.01in})\hspace*{-.01in}=\hspace*{-.01in}0$. For fixed $m_k$ with \small$\displaystyle\sum_{k=1}^K m_k\hspace*{-.03in}=\hspace*{-.03in}m$\normalsize, there are ${m\choose {m_1,\dotsc,m_K}}$ possibilities for $\bfm F_y$ leading to $ H(\bfm F_{\hspace*{-.01in}y}|m_{\hspace*{-.01in}1},\dotsc, m_{\hspace*{-.01in}K})\hspace*{-.01in}\leq\hspace*{-.01in} \log{m\choose {m_1,\dotsc,m_K}}$. It follows from the inequality (see Appendix~\ref{app-inequality-logM}) 
\begin{equation}\label{eq-inequality-logM}
\log{m\choose {m_1,\dotsc,m_K}}\leq m\log(m)-\sum_{k=1}^K m_k\log(m_k),
\end{equation} 
that \mbox{$\displaystyle H\hspace*{-.01in}(\hspace*{-.01in}\bfm F_{\hspace*{-.01in}y}| m_{\hspace*{-.01in}1},\dotsc, m_{\hspace*{-.01in}K\hspace*{-.01in}})\hspace*{-.015in} \leq\hspace*{-.015in} m\hspace*{-.01in}\log(m)\hspace*{-.01in}-\hspace*{-.01in}\sum_{k=1}^K m_k\log(m_k)$}.~Since $\displaystyle {g([m_1,\dotsc,m_k])\hspace*{-.02in}=\hspace*{-.025in}\left(\hspace*{-.01in}\sum_{k=1}^K m_k\hspace*{-.01in}\right)\hspace*{-.02in}\log\hspace*{-.01in}\left(\hspace*{-.01in}\sum_{k=1}^K m_k\hspace*{-.01in}\right)\hspace*{-.03in}-\hspace*{-.03in}\sum_{k=1}^K m_k\log(m_k)}$ is a concave function of $[m_1,\dotsc,m_K]$ (see Appendix~\ref{app_concave}), employing the Jensen's inequality yields 
\begin{equation}\label{eq-I_F-UB}
I_F \hspace*{-.04in}\leq\hspace*{-.04in} \left(\hspace*{-.01in}\sum_{k=1}^K \hspace*{-.03in} E\{M_k\}\hspace*{-.03in}\right)\hspace*{-.03in}\log\hspace*{-.02in}\left(\hspace*{-.01in}\sum_{k=1}^K \hspace*{-.03in} E\{M_k\}\hspace*{-.03in}\right)
\hspace*{-.02in}-\hspace*{-.02in}\sum_{k=1}^K \hspace*{-.02in}E\{M_k\}\hspace*{-.02in}\log(\hspace*{-.02in}E\{M_k\}\hspace*{-.02in}).\nonumber
\end{equation}
On the other hand, due to the fact that ${\cal{C}}_k$ are i.i.d. binary input deletion channels, we have $E\{M_k\}=N(1-d)\alpha_k$ where $\alpha_k$'s depend on the input distribution $P(\bfm X)$ and $\sum_{k=1}^K\alpha_k=1$. Hence, we obtain
\begin{align}\label{eq-I_F-UB}
I_F \leq & N(1-d)\left(\log\left(N(1-d)\right)-\sum_{k=1}^K\alpha_k\log\left(N(1-d)\alpha_k\right)\right)\nonumber\\
= & -N(1-d)\sum_{k=1}^K \alpha_k\log{\alpha_k}=N(1-d)H(\alpha_1,\dotsc,\alpha_K)\nonumber\\
\leq & \ N(1-d)\log(K),
\end{align}
which concludes the proof.
\end{IEEEproof}

\vspace*{-.02in}\subsection{Proof of Theorem~\ref{thm-2K-2}}\vspace*{-.02in}
Substituting the results of Lemmas~\ref{lemma-Ik} and \ref{lemma-IF} into \eqref{eq-I_2K}, we obtain\vspace*{-.1in}
\begin{align}
I(\bfm X;\bfm Y)\leq&\ E_{N_1,\dotsc,N_K}\left\{\sum_{k=1}^K  N_k\right\} C_2(d)+ 2K \log(N+1)\nonumber\\
&+N(1-d)\log(K)\nonumber\\
=&\ N C_2(d)+ 2K \log(N+1)+N(1-d)\log(K),\nonumber
\end{align}
where we have used the fact that $\sum_{k=1}^K N_k=N$ independent of the input distribution $P(\bfm X)$. Since the above inequality holds for any input distribution $P(\bfm X)$ and any value of $N$, we can write
\begin{eqnarray}
C_{2K}(d)&=&\lim _{N\to\infty} \max_{P(\bfm X)}\frac{1}{N} I(\bfm X;\bfm Y)\nonumber\\
&\leq & C_2(d)+(1-d)\log(K),\nonumber
\end{eqnarray}
which concludes the proof of Theorem~\ref{thm-2K-2}.\hfill$\Box$

\vspace*{-.05in}\section{Some Implications}\label{sec-Numerical-Examples}

As stated earlier, a trivial upper bound on the capacity of the $2K$-ary deletion channel is given by $(1-d)\log(2K)$ which is the capacity of the $2K$-ary erasure channel. We have shown in the previous section that by substituting any upper bound on the capacity of the binary deletion channel into~\eqref{eq-C2K-C2}, an upper bound on the $2K$-ary deletion channel capacity results. Obviously, by employing $C_2(d)\leq 1-d$, which is the trivial upper bound on the binary deletion channel capacity, the erasure channel upper bound on the $2K$-ary deletion channel capacity is obtained. Therefore, any upper bound tighter than $1-d$ on the binary deletion channel capacity gives an upper bound tighter than $\log(2K)(1-d)$ on the $2K$-ary deletion channel capacity. The amount of improvement is $1-d-C_2^{UB}(d)$, where $C_2^{UB}$ denotes the upper bound on the binary deletion channel capacity. 

As it is shown in~\cite{mercier2012}, $(1-d)\log(2K)-1\leq C_{2K}(d)\leq (1-d)\log(2K)$, where the lower bound is implied from~\eqref{eq-LB-diggavi-iid}, therefore the existing trivial upper and lower bounds are tight enough for asymptotically large values of $K$, and i.i.d. distributed input sequences are sufficient to achieve the capacity. However, the importance of the result in Theorem~\ref{thm-2K-2} is for moderate values of $K$, where the amount of improvement in closing the gap between the existing upper and lower bounds is significant. 

\begin{figure}[t]
\centerline{
  \hbox{
 \resizebox{85mm}{!}
    {\includegraphics[trim=7mm 4mm 10mm 5mm,clip,  width=.5\textwidth]{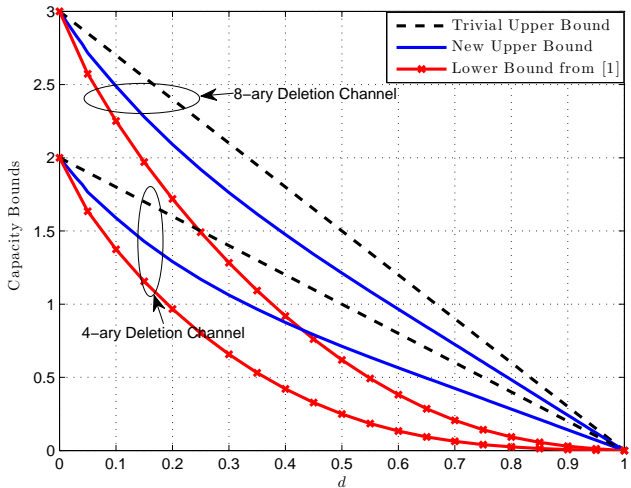}}
  }
}
\vspace*{-.1in}\caption{\hspace*{-.18in} Comparison among the new upper bound~\eqref{eq-C2K-C2}, the lower bound~\eqref{eq-LB-diggavi-Markov} and the trivial erasure channel upper bound for the $4$-ary and $8$-ary deletion channels.\vspace*{-.1in}}
\label{fig-upper-bound-lower-bound}
\end{figure}

To demonstrate the improvement over the trivial erasure channel upper bound, we compare the upper bound ${C_{2K}(d)\leq C_2^{UB}(d)+(1-d)\log(K)}$ with the erasure channel upper bound $\log(2K)(1-d)$ and the tightest existing lower bound~\eqref{eq-LB-diggavi-Markov} (from~\cite{diggavi2006information}) in Fig.~\ref{fig-upper-bound-lower-bound} for $4$-ary and $8$-ary deletion channels. Here we utilize the binary deletion channel capacity upper bounds $C_2^{UB}(d)$ in~\cite{dario, IT-quasi}, where for $d\leq 0.65$ we use the results in~\cite[Table III]{dario} and for $d\geq 0.65$ we use the upper bound $C_2(d)\leq 0.4143(1-d)$ given in~\cite{IT-quasi}.

\begin{figure}[t]
\centerline{
  \hbox{
 \resizebox{85mm}{!}
    {\includegraphics[trim=7mm 4mm 13mm 7mm,clip,  width=.8\textwidth]{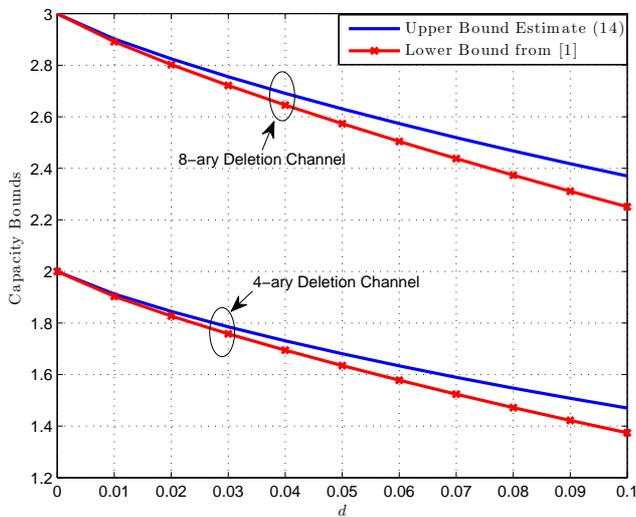}}
  }
}
\vspace*{-.1in}\caption{Comparison between the upper bound~\eqref{eq-LB-K-ary-small-d} (ignoring the $O(d^{3-\epsilon})$ term) and the lower bound~\eqref{eq-LB-diggavi-Markov}.\vspace*{-.3in}}
\label{fig-upper-bound-estimate}
\end{figure}

Another implication of the result in Theorem~\ref{thm-2K-2} is in studying the asymptotic behavior of the $2K$-ary deletion channel capacity for $d\to 0$. It is shown in~\cite{kanoria-arxiv} that\vspace*{-.02in} 
\begin{equation}\label{eq-bdeletion-capacity-expansion}
C_2(d)=1+d\log(d)-A_1d+A_2d^2+O(d^{3-\epsilon}),
\end{equation} for small $d$ and any $\epsilon>0$ with $A_1\approx 1.15416377
$, ${A_2\approx 1.78628364}$ and $O(.)$ denoting the standard Landau (big-O) notation. Employing this result in~\eqref{eq-C2K-C2}, leads to an upper bound expansion for small values of $d$ as\vspace*{-.03in} 
\begin{align}\label{eq-LB-K-ary-small-d}
C_{2K}(d)\leq\ &1+d\log(d)-(A_1+\log(K))d+ A_2 d^2+ \log(K)\nonumber\\
&+O(d^{3-\epsilon}).
\end{align}
In Fig.~\ref{fig-upper-bound-estimate}, we compare the above upper bound (by ignoring the $O(d^{3-\epsilon})$ term) which serves as an estimate, with the lower bound~\eqref{eq-LB-diggavi-Markov} for $d\leq 0.1$. We observe that by employing the capacity expansion~\eqref{eq-bdeletion-capacity-expansion} in~\eqref{eq-C2K-C2}, a good characterization for the asymptotic behavior of the $2K$-ary deletion channel capacity is obtained as $d\rightarrow 0$.

\vspace*{-.08in}\section{Conclusions}\label{sec-Conclusions}\vspace*{-.05in}

We have derived the first non-trivial upper bound on the $2K$-ary deletion channel capacity. We first considered the $2K$-ary deletion channel as a parallel concatenation of $K$ independent binary deletion channels, all with the same deletion probability. We then related the capacity of the original channel to that of the binary deletion channel. By doing so we obtained an upper bound on the capacity of the $2K$-ary deletion channel in terms of the capacity of the binary deletion channel and as a result any upper bound on the capacity of the binary deletion channel. The provided upper bound results in tighter upper bounds on the $K$-ary deletion channel capacity than the trivial erasure channel upper bound for the entire range of deletion probabilities. 

\appendices

\vspace*{-.09in}\section{Proof of Inequality~\eqref{eq-inequality-logM}}\label{app-inequality-logM}\vspace*{-.05in}
It follows from the inequality $\log{m\choose {m_1}}\leq {mH_b(\frac{m_1}{m})}=m\log\left(m\right)-m_1\log\left(m_1\right)-\left(m-m_1\right)\log\left(m-m_1\right)$ given in~\cite[p.~353]{cover} that\vspace*{-.04in}
\begin{align}
\log&{m\choose {m_1,\dotsc, m_K}}=\sum_{j=1}^{K-1}\log {{m-\sum_{k=1}^{j-1}m_k}\choose {m_{j}}}\nonumber\\
& \leq \sum_{j=1}^{K-1} \left(m-\sum_{k=1}^{j-1}m_k\right)\log{\left(m-\sum_{k=1}^{j-1}m_k\right)}-m_j\log{m_j}\nonumber\\
&\hspace*{.1in} -\sum_{j=1}^{K-1}\left(m-\sum_{k=1}^{j}m_k\right)\log{\left(m-\sum_{k=1}^{j}m_k\right)}\nonumber\vspace*{-.04in}\\
&=m\log(m)-\sum_{k=1}^K m_k\log(m_k).\nonumber 
\end{align}

\vspace*{-.1in}
\section{Concavity of $g([m_1,\dotsc,m_k])$}\label{app_concave}
For the Hessian of $g([m_1,\dotsc,m_k])$, we have
\begin{equation}
\nabla^2 g([m_1,\dotsc, m_k])= \frac{1}{\sum_{k=1}^K m_k} \bfm 1\bfm 1^T-diag\left\{\frac{1}{m_1},\dotsc,\frac{1}{m_K}\right\},\vspace*{-.04in}\nonumber
\end{equation}
where $\bfm 1$ is an all one vector of length $K$, i.e., $\bfm 1=[1,\dotsc,1]^T$, and $diag\left\{\frac{1}{m_1},\dotsc,\frac{1}{m_K}\right\}$ denotes a diagonal matrix whose $k$-th diagonal element is $\frac{1}{m_k}$. Furthermore, by defining $\bfm a=[a_1,\dotsc,a_K]$, we can write\vspace*{-.04in}
\begin{align}\displaystyle
\bfm a \nabla^2 g\bfm a^T = & \frac{(\sum_{k=1}^K a_k)^2}{\sum_{k=1}^K m_k}-\sum_{k=1}^K\frac{a_k^2}{m_k}\nonumber\\
=& \frac{1}{\sum_{k=1}^K m_k}\bigg(\sum_{k=1}^K a_k^2 +2\sum_{k=1}^{K-1}\sum_{j=k+1}^K a_ka_j\nonumber\\
&\quad\quad\quad \quad\quad -\sum_{k=1}^K a_k^2 -\sum_{k=1}^K \frac{\sum_{j\neq k}m_j}{m_k} a_k^2 \bigg)\nonumber\\
=&\frac{1}{\sum_{k=1}^K m_k}\sum_{k=1}^{K-1}\sum_{j=k+1}^K\left( 2a_ka_j-\frac{m_j}{m_k}a_k^2-\frac{m_k}{m_j}a_j^2\right)\nonumber\\
=&\frac{-1}{\sum_{k=1}^K m_k} \sum_{k=1}^{K-1}\sum_{j=k+1}^K \frac{m_j}{m_k}(a_k-\frac{m_k}{m_j}a_j)^2,\nonumber
\end{align}
which is negative for all $m_k,\ m_j>0$. Therefore, $\nabla^2 g([m_1,\dotsc,m_k])$ is a negative semi-definite matrix and as a result $g([m_1,\dotsc,m_k])$ is a concave function of $[m_1,\dotsc,m_k]$.\vspace*{-.03in}

\section*{Acknowledgment}\vspace*{-.01in}
The authors are supported by the National Science Foundation under the contract {NSF}-{TF} 0830611. Tolga M. Duman is also supported by the EC Marie Curie Career Integration Grant PCIG12-GA-2012-334213.\vspace*{-.03in}

\bibliographystyle{IEEEtran}
\footnotesize
\bibliography{myrefs}

\end{document}